**Geant4 based Dosimetry Evaluation for Gamma Knife using Different Phantom Materials**

Özlem DAĞLI[1], Erkan BOSTANCI[2], Ömer Hakan EMMEZ[3], Gökhan KURT[3], Fatih EKİNCİ[4] and Mehmet Serdar GÜZEL[2]

1 Faculty of Medicine, Department of Brain and Neurosurgery, Gazi University, Ankara, Turkey
E-mail Address: ozlemdagli@gazi.edu.tr
2. Department of Computer Engineering, Ankara University, Ankara, TURKEY
E-mail Address: ebostanci@ankara.edu.tr, mguzel@ankara.edu.tr
3. Faculty of Medicine, Department of Brain and Neurosurgery, Gazi University, Ankara, Turkey
Email Address: hakanemmez@gmail.com, gkurtmd@gmail.com
4. Gazi University, Physics Department, Ankara, TURKEY
E-mail Address: fatih.ekinci2@gazi.edu.tr

Abstract. This study analyses the dose difference for a variety of phantom materials that can be used for Leksell Gamma Knife. These materials have properties that are very similar to the human tissue not including the skull bone. Geant4 was employed in the analysis the dose distributions for collimator helmet sizes of 4mm and 8mm. The phantom had a radius of 80mm. Water, brain, PMMA (Poly-methyl methacrylate) and polystyrene were used as the material types. Results showed no considerable differences for radiation dosimetries depending on the material types. In addition, the polystyrene and PMMA (Poly-methyl methacrylate) phantom are also suitable for measuring the dose profiles of the Gamma Knife unit.

## 1. Introduction

Gamma Knife device developed in 1950s, was first presented in 1967 using 179 Co-60 sources. Early Acoustic Neuroma patient, was treated by Leksell in1969. In 1975 the second gamma knife device was established under the Karolinska Institute and Brain Surgery service. The third and fourth units were used in Buenos Aires/Argentina and Sheffield/England. It was initially applied to functional neurosurgery patients and then to some benign tumors and small sized malignant tumors in the 1980s. The primary 201 Cobalt-60 sources Gamma Knife was founded in the USA in 1987 and later sophisticated the Model C followed by 192 Cobalt-60 sources Leksell Gamma Knife Perfexion. Gamma Knife Icon is the sixth and recent generation of the Leksell Gamma Knife technology [1].

Radiosurgery, a conformational treatment method, involves directing target bunches of beams from a number of different angles, resulting in rapid dose droping in normal tissues outside the target, while high doses are achieved in the confluence of the rays.

Leksell Gamma Knife radiosurgery involves no institutionary surgical incisions for brain surgery [2]. The principle of Gamma Knife radiosurgery is incomplex. GammaPlan study a tissue equivalent material, with an attenuation coefficient μ =0.0063 mm-1 at the energy 1.25 MeV, in all calculations without the existing of a skull bone. Furthermore, in routine quality assurance programs of the Gamma Knife unit, a spherical polystyrene phantom is study to give dose distributions with the opening of all 201 60Co sources. This phantom may not be completely tissue equivalent. Therefore, compatibility of these dose distributions is complicated [3]. In the present study, we used the Monte

Carlo method Geant4 to calculate the radial dose distributions from a single radiation beam of 4 mm and 8 mm collimator helmet in variant phantom materials.

The source of radiation received in radiotherapy can be particle-based as well as gamma [4,5]. Different sources of radiation cause different reactions within the target [6]. At the same time, the increasing importance of radiation-related applications in our lives has increased the importance of measuring the amount of radiation received by living organisms or any material exposed to radiation [7].

Geant 4 is a Monte Carlo simulation software with a large library of physics, including tools that can simulate absolutely the interaction of the particles with the target. Geant name, "it was created using the geometry and Tracking words. While the major goal of the software growth is high energy physics experiment simulations, it is so used in many areas such as nuclear physics, medical and astrophysics today owing to the success and requirements of detector simulations. There are several ways to get a Geant4 simulation for a particular problem. The easiest is to use a ready-made application or tool that provides the essential qualifications to create an installation or detector adapted to the field of application and to measure the observables. At the basis of Geant4 software, a category scheme is needed for the designed detector structure to be able to determine both geometric and physical phenomena and read the results. Starting from the root of the scheme and moving upwards, the essential structures and procedures for the simulation are performed, so that a subaltern structure of the simulation operation in a certain order and order is created [8].

In this study, dosimetry calculations were made for different phantom materials that can be used for Leksell Gamma Knife device. Water, brain, PMMA and polystyrene were used as the phantom material types. Solid water, brain, PMMA and polystyrene phantom taken into consideration and the design of the skull and device source was made with the Geant-4 computer-based program, and the target doses were determined.

The rest of the paper is structured as follows: Section 2 describes the material and the method employed in our study, followed by Section 3 where findings are presented and discussed. Finally, the paper is concluded in Section 4.

## 2. Material and Method

In the study, a Gamma Knife device, brain phantom was simulated using Geant4. The Gamma Knife Device (GK) used was the 4C model with 201 sources. Phantoms are materials that are equivalent to human tissue and which are used to examine dose distributions in tissue.

In the present study, we used the Monte Carlo method Geant4 to calculate the radial dose distributions from a single radiation beam of 8 mm and 14 mm collimator helmet in different phantom materials. The material compositions of these phantom obtained from ICRP [9] are shown in Table 1.

Table 1 Phantom materials used in the analysis

| Component Name | Chemical Formula | Density |
|---|---|---|
| PMMA | (C5O2H8)n | 1,18 g/ml |
| Polystyrene | (C8H8)n | 0.96–1.05 g/ml |

In this study, a 80 mm radius phantom was used. As a phantom material are selected solid water, brain, PMMA (poly-methyl methacrylate) and polystyrene.

Geant4 is a modern Monte Carlo simulation program that emerged in 1993 with the work of scientists at CERN (European Organization for Nuclear Research), which can simulate particles interacting with matter. The name Geant is derived from the words "GEometry ANd Tracking" [10].

In Geant4 application, firstly, the preparation of geometry such as materials to be simulated, volumes and locations; description of related physics such as particles, physical processes, models and production threshold energy; formation mechanism of primary particles; display of prepared geometry and particle traces; adding user User Interface (UI) commands; During the simulation, necessary information must be collected [11].

For the simulation application of the Gamma Knife device, a new modular design was made by using the documentation of the existing system and taking its technical features. The aim here is to define the gamma beam profile and to reveal the most appropriate and useful structure for obtaining beam profile by using existing irradiation-collimator materials.

In the Geant4 simulation program, while preparing the simulation software of the Gamma Knife system, four basic steps were taken into consideration: 1) Defining the general structure, 2) creating the physical geometry, 3) defining the physics events and 4) simulation process and calculations.

3D models of Gamma Knife design were prepared and preliminary graphic models for simulation were created in computer environment. The 3D designs of the source are shown in Figure 1 below [12].

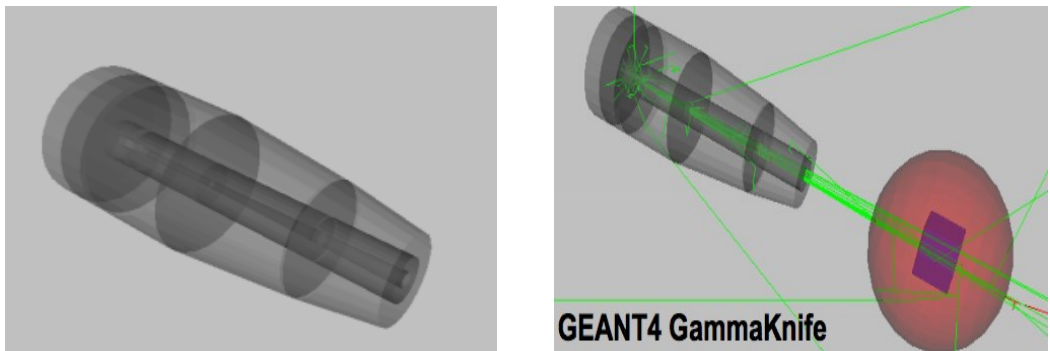

**Figure 1.** 3D resource design with Geant4 [13]

The shapes forming three basic geometric structures as World, Target and Tracker are defined with codes in Geant4 software. According to this definition, the world; A room with an indoor environment of 400x400 cm² air represents the position in the modeling and the model of the environment where the experiment will take place.

Target was defined as 90x90x90 mm voxel detector plate. Other materials were also defined as similarly. These definitions; While creating the basic geometric structure, the collimator head and dimensions that ensure the exact fit of the model are defined. Geometric definitions of other objects are made similarly [14].

In the last step of the simulation study; It includes simulation processes and physical calculations in a way that is exactly appropriate for the model. The Gamma knife device consists of a hemispherical iron-coated unit containing 201 co-60 sources. Gamma rays come from different directions and focus on the target.

Calculations were made for 4, 8 mm collimators. The stored energy was calculated at the end of the simulation in voxels divided into small 90x90x90 mm cubes using scoring mesh. In the modeling study, firstly, instead of 201, the gamma ray originating from a single source was taken as a reference and calculations were made for 201 sources at different angles.

While modeling, the spherical water phantom with a radius of 80 mm, the brain phantom taken from ICRU data, PMMA (Poly-methyl methacrylate) and polystyrene phantom were used in dosimetric measurements (ICRU). The distance between patient source is 401 mm. Two gamma rays of 1.17 MeV and 1.33 MeV are released from the Co-60 decay. The stored energy was plotted by normalizing the dose with the data taken from the simulations.

### 3. Results and Discussion

The simulation results were used to analyse the differences between varying embolization materials. The findings are described in the following.

The absorbed doses were compared in Table 2. The difference between the brain and the water phantom was observed as 12.5% and 7.9% for collimators with diameters of 4mm and 8 mm, respectively. The difference between the brain and the PMMA phantom was found as 12.6% when using a 4mm diameter collimator while it was 1.4% for the 8mm diameter collimator. The difference between the brain and the Polystyrene phantom was observed as 6.2% and 3.6% for collimators with diameters of 4mm and 8 mm, respectively.

**Table 2.** Comparison of absorbed doses for different phantoms materials

| Collimator helmet size | Mean peak dose Phantom (a) | Mean peak dose Phantom (b) | % difference $|(a-b)|\times 100/a$ |
|---|---|---|---|
| 4 mm | Brain 0,8420 | Water 0.9473 | 12.5 |
|  |  | PMMA 0.9481 | 12.6 |
|  |  | Polystyrene 0.8945 | 6.2 |
| 8 mm | Brain 0,9268 | Water 0.8530 | 7.9 |
|  |  | PMMA 0.9399 | 1.4 |
|  |  | Polystyrene 0.8928 | 3.6 |

When Figures 2 and 3 are examined, it can be seen that the normalized dose curve expands as the collimator diameter increases. That is, as the collimator diameter increased, normal tissue received more doses. In addition, another observation is that as the collimator diameter increases, the margin of error decreases.

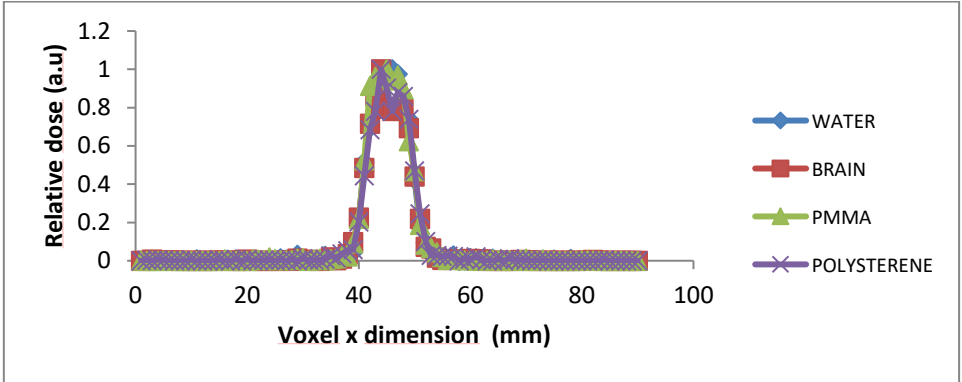

**Figure 2.** A comparison of radial doses in different phantom materials from the 4 mm collimator helmet of the Leksell Gamma Knife

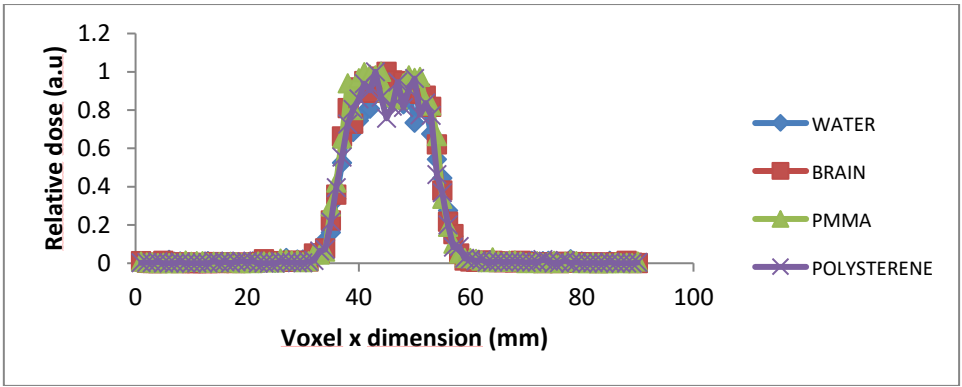

**Figure 3.** A comparison of radial doses in different phantom materials from the 8 mm collimator helmet of the Leksell Gamma Knife

Table 2. shows the results of the t-test performed to find out whether there are significant differences in the dose difference computed for different embolization materials. The simulations were run 10 times and the dose accumulations are compared. From the results, one can conclude that there are significant differences between water and brain while the differences of PMMA and Polystyrene were not found to be statistically significant.

**Table 2** Statistical evaluation of dose differences of phantom materials with brain t-test

|  | Water | PMMA | Polystyrene |
|---|---|---|---|
| t Stat | 4.03118 | -0.27768 | 1.120481 |
| P(T<=t) one-tailed | 0.000713 | 0.392812 | 0.141391 |
| t critical , one tailed | 1.770933 | 1.770933 | 1.770933 |
| P(T<=t) two-tailed | 0.001426 | 0.785624 | 0.282782 |
| t critical , two tailed | 2.160369 | 2.160369 | 2.160369 |

## 4. Conclusion

The correct determination of the radiation dose given to the patient is critically important. Therefore, phantoms are used that are equivalent to human tissue and examine the dose distribution in the tissue.

Our results show that the differences between the dose values using Geant4 for solid water phantom, brain, PMMA (Poly-methyl methacrylate) and polystyrene are remarkable. The percentage difference of Table 1 indicates some differences; however, these differences were not found to be statistically significant when the t-test was applied. Furthermore, these differences in statistical terms are not larger than the acceptable range (< 3%) prescribed by the stereotactic radiosurgery. Results show different radiation dosimetries depending on the material types. In addition, the polystyrene and PMMA phantom are also suitable for measuring the dose profiles of the Gamma Knife unit [15].

Dose distribution in Leksell Gamma Plan (LGP) was calculated based on a homogeneous phantom. Different results were found when materials of different densities were included.

Comparing the obtained results encourages use using Monte Carlo Geant4 simulations to investigate further problems related to SRS [16].